\documentstyle[12pt]{article}
\newcommand{\beq}{\begin{equation}}
\newcommand{\eeq}{\end{equation}}
\newcommand{\bea}{\begin{eqnarray}}
\newcommand{\eea}{\end{eqnarray}}

\input epsf
\begin{document} 

\hfill \vbox{\hbox{UCLA/01/TEP/14}} 
\begin{center}{\Large\bf Light fermion mass generation  
in dynamical symmetry breaking}\\[2cm] 
{\bf E. T. Tomboulis}\footnote{Research supported by 
NSF grant NSF-PHY 9819686}\\
{\em Department of Physics, UCLA, Los Angeles, 
CA 90095-1547}\\
{\sf e-mail: tombouli@physics.ucla.edu}
\end{center}
\vspace{1cm}

\begin{center}{\Large\bf Abstract} 
\end{center}
We reconsider the question of mass generation for 
fermions coupled to a set of gauge bosons (in 
particular, the electroweak gauge bosons) when the latter get 
their masses through the Goldstone bosons originating 
in a simple (i.e. not extended) technicolor sector. The  
fermion global chiral symmetries are broken by including 
four-fermion interactions. We find that the system can be 
nonperturbatively unstable under fermion mass fluctuations 
driving the formation of an effective coupling 
between the technigoldstone bosons and the ordinary fermions.   
Minimization of an effective action for the 
corresponding composite operators describes then  
dynamical generation of light fermion masses $\sim M \exp(-k/g^2)$, 
where $M$ is some cutoff mass.

\vfill
\pagebreak 

\section{Introduction}
    
Electroweak dynamical symmetry breaking (DSB) provides a  
natural and attractive mechanism for generating the $W$ and $Z$ 
masses. It has proven much more difficult, however, to 
satisfactorily account for  
the quark and lepton masses within such a framework. 
Extended technicolor, walking technicolor, top condensation,  
and top-color assisted technicolor are among the various 
proposals that have been investigated. (For reviews and references,   
see \cite{C}.)

In this paper we reconsider the question of fermion 
mass generation in a framework employing only  
simple (i.e. not extended) technicolor.    
Specifically, we investigate the following question. 
Consider the system consisting of 
a simple technicolor sector, electroweak interactions, and the 
ordinary quarks and leptons.  
In addition, postulate four-fermi interactions among the 
ordinary fermions that explicitly break their global chiral symmetries. 
Thus, restricting to one fermion family only,     
introduce the interactions \cite{MTY}: 
\beq 
{\cal L}_{4f} = {4\pi^2\over N_c\Lambda^2}\Bigg[ \,
G_1\,(\bar{q}_L^i q_R^j)(\bar{q}_R^j q_L^i) + 
G_3\,(\bar{q}_L^i q_R^j)\,\tau^3_{jk}(\bar{q}_R^k q_L^i) 
+ G_2\, (\bar{q}_L^i q_R^j)\,\epsilon_{ik}\epsilon_{jl}
(\bar{q}_L^k q_R^l) +\mbox{h.c.}\Bigg] \label{4f}
\eeq 
where sum over color indices is understood, $i,j,k,l$ are 
isospin indices, and $\Lambda$ some UV cutoff. 
(\ref{4f}) reduces the weak symmetry group to just 
$SU(2)_L\times U(1)_Y$.\footnote{Indeed, the symmetry of the three 
terms is $SU(2)_L\times SU(2)_R\times U(1)_V\times U(1)_A$, 
$SU(2)_L\times U(1)_Y\times U(1)_V\times U(1)_A$, and 
$SU(2)_L\times SU(2)_R\times U(1)_V$, respectively.} Recall 
that in the standard Higgs model this is a function fulfilled 
by the Yukawa couplings.    
Note that for sufficiently large values of the $G_1$ coupling, 
this four-fermion interaction can induce dynamic chiral symmetry 
breaking and mass generation (NJL model), as in fact is assumed 
in top condensation schemes \cite{C}. Here we will always 
assume that four-fermion 
couplings in (\ref{4f}) are below their critical value 
for inducing any mass generation effects solely by themselves.     
 
The electroweak gauge bosons are assumed to acquire mass 
through the Goldstone bosons associated with CSB in the 
technicolor sector. We then ask 
whether in this system mass generation for the
ordinary fermions can also occur.  

Now since, with only the interactions specified above present,  
the ordinary fermions 
can communicate with the techniquark sector only through 
the electroweak gauge bosons, and gauge interaction vertices 
preserve chirality, it is clear that this cannot happen to 
any finite order in  the couplings. The question is whether 
it can happen nonperturbatively. The conventional answer to this 
would appear to be negative: the weak interactions are `too weak' 
to produce such dynamical mass generation. This, however, is actually 
a spurious argument. The weakness of the weak interactions could be  
no more relevant here than it is in the generation of the $W$ and $Z$ 
masses. It is the strong technicolor interactions that 
produce the necessary Goldstone bosons, and  the relevant question is 
whether an effective Yukawa coupling can form nonperturbatively 
between these Godstone bosons and the quarks. What determines this, in 
physical terms, is whether the coupled system is unstable under mass 
fluctuations. If it is, then even very weak interactions can 
drive the instability (even though the instability cannot be 
seen to any finite order in perturbation theory).

The same question arises more generally when   
the standard electroweak gauge bosons are replaced by some other set 
of gauge bosons at a different, perhaps much higher, mass scale.   
In fact it is in this form that the question would more likely be   
pertinent to mass generation for all three fermion generations, 
as well as more general use of DSB in the construction of models.  
In the following we consider only the case of the    
standard electroweak interactions with one fermion family, 
and the simplest QCD-like technicolor sector;  
but the same analysis 
applies in the more general context. This analysis 
leads to a definition of an effective action, and 
hence, by minimization, to a set of self-consistent 
equations for dynamically generated effective Yukawa vertices. 
Unfortunately, consideration of these equations in general entails   
very considerable computational complexity, but the physical 
context is transparent. 
The trivial (perturbative) solution is always a solution,  
but one finds that, depending on the model, 
a nontrivial solution may also exist. 
From the structure of the 
self-consistent equations, it follows that 
a nontrivial solution, when it exists, generically describes 
dynamical generation of fermion masses $\sim M \exp(-k/g^2)$, 
where $k$ depends on the couplings $G_i$ in (\ref{4f}), and 
$M$ is some (cutoff) mass. In the case of the electroweak 
interactions treated below this gives very small fermion masses  
$\sim m_W \exp(-{\rm k}/g^2)$.     
  
\section{\hbox{\bf Preliminaries}}
We consider the minimal technicolor theory with technicolor 
gauge group $G_{TC}$, and two flavors of 
fundamental representation massless 
techniquarks $Q =(U,\;D)$. (It is important for our purposes to 
work within a model that allows a certain amount of computation. 
$G_{TC}=SU(2)$ may in fact be the only 
experimentally still viable QCD-like simple technicolor model.)   
The global chiral group is then $SU(2)_L\times SU(2)_R$, and is 
spontaneously broken by the strong technicolor interaction 
to diagonal $SU(2)$ resulting in mass $M$ for the technifermions, 
and a triplet of Goldstone bosons (the technipions) $\phi^a$. 
The corresponding broken chiral generators are given by $\gamma^5\tau^i$, 
$i=1,2,3$, where $\tau^i$ are the Pauli matrices. 
The coupling of the Nambu-Goldstone bosons to the technifermions  
is, for momenta below a cutoff $\Lambda \sim M$, adequately 
represented in a simple NJL model (for review and references, 
see e.g. \cite{dR}),  
or an equivalent linear sigma model effective description, giving 
\beq
\begin{minipage}{3cm}
\epsfysize=2cm\epsfbox{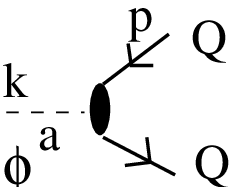}
\end{minipage} \quad = \gamma^5\tau^a G\,P(p+k,p) 
\label{gold} 
\eeq
with (dimensionless) wave-function 
$P(p+k,p) = 1 + O(k/M) $. For momenta above $\Lambda$, $P(p,k)$ 
decays rapidly to zero -- the detailed UV behavior will in fact be 
irrelevant for physics at scales well below $M$. The effective 
coupling $G$ is related to the technipion decay constant $F$ 
by the GT relation $M=FG$. The triplet of pions is accompanied by 
a massive bound state, the sigma or real higgs scalar, of mass $2M$.

\begin{figure}[htb]
{\hfill\epsfysize=2cm\epsfbox{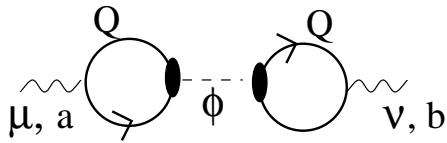}\hfill}
\caption[vacp]{\label{tcvacp}Schwinger mechanism} 
\end{figure}

For application to standard electroweak theory, the techniquarks form 
an electroweak left-handed doublet 
$Q_L= (\,U_L\; D_L\,)$, and right-handed singlets $U_R$, $D_R$. 
Anomalies are cancelled by assigning  
hypercharge $0$ to $Q_L$, and $\pm1/2$ to $U_R$, $D_R$, respectively. 
As it is well-known, the electroweak gauge bosons then acquire mass through 
the pole in their polarization tensor (Schwinger mechanism) 
generated by the coupling of the Goldstone bosons (figure \ref{tcvacp}). 
The resulting mass matrix   
reproduces the familiar electroweak gauge boson mass matrix with    
$m_W=gF/2$.   

It should be noted that the pole contribution represented in 
figure \ref{tcvacp}
gives only the $k^\mu k^\nu$ part of the   
polarization tensor $\Pi^{\mu\nu}_{ab}(k)=
(k^2 g_{\mu\nu} - k^\mu k^\nu)\Pi_{ab}(k^2)$.  
It is not easy to identify    
directly in terms of diagrammatic contributions, such as in 
figure \ref{tcvacp}, the accompanying $g_{\mu\nu}$-part 
which must be there because of gauge 
invariance. This is typical of DSB computations where one often 
must rely on general, in principle non-perturbative, constraints 
to trace the symmetry breaking effects. In the present case,   
the transverse polarization tensor is related to the 
3-gauge boson proper vertex by a Ward identity, which in 
the zero momentum transfer limit becomes 
\beq 
\lim_{q\to 0} q^\mu\,\Gamma^{\mu\kappa\lambda}_{abc}(q,k,-k-q) = 
-i(k^2 g^{\kappa\lambda} - k^\kappa k^\lambda )\,[\,(T^a - B^a), 
\Pi(k^2)\,]_{\ \atop \scriptstyle bc}
.\label{WI1a} 
\eeq   
Here $T^a$ are the adjoint generators of $SU(2)\times U(1)$.  
(Coupling constants are included in the definition of generators.) The 
matrix $B^a$, also proportional to $T^a$, involves the 
FP ghosts and is actually irrelevant for the development below. 
The pole contribution $\Pi_{ab}(k^2){ \ \atop{\longrightarrow\atop k^2\to 0}}
\mu^2_{ab}/k^2$ then satisfies 
\beq
\lim_{q,k \to 0} q^\mu\,\Gamma^{\mu\kappa\lambda}_{abc}(q,k,-k-q) = 
(g^{\kappa\lambda} - {k^\kappa k^\lambda\over k^2} )\,[\,T^a , 
\mu^2\,]_{\ \atop \scriptstyle bc} \, .\label{WI1b}
\eeq 
A Goldstone pole with nonvanishing residue 
in the 3-gauge boson proper vertex then implies a nonvanishing 
symmetry-violating  mass matrix such that the commutator 
$[\,T^a, \mu^2\,]$ on the r.h.s. does not vanish. 
The occurrence of such poles follows from the 
existence of nonvanishing effective $\phi^{\pm}W^{\mp}V$ couplings, 
($V=\gamma, Z$), as is easily verified by explicit computation 
using (\ref{gold}) and the GT relation. The resulting residues 
on the l.h.s. of (\ref{WI1b}) precisely match the commutator 
on the r.h.s. computed with the 
mass matrix surmised from figure \ref{tcvacp}. 

\section{How the fermions can get mass}

Consider now the coupling of the ordinary quarks and leptons. 
It suffices to consider a singlet doublet $q=(u,d)$.   
The gauge boson-quark-quark proper vertex is related to the 
inverse quark propagator $iS^{-1}(p) = \slash{p} - \Sigma(p)$ 
by the non-Abelian version of the original QED WI. In the 
zero momentum transfer limit one has   
\beq
\lim_{q\to 0} q_\mu \Gamma^{\mu a}_{ij}(p+q,p) = 
 -i\Sigma_{il}(p)\,[\,t^a_{lj} - B^a_{lj}(p)\,] 
+ i \gamma^0\,[\,t^a_{il}- 
B^a_{il}(p)\,]\,\gamma^0 \Sigma_{lj}(p) .\label{WI2a} 
\eeq  
The $t^a$'s denote the generators in the fermion representation. 
Again, the quantity $B^a_{ij}$ involves the FP ghosts 
and need not be given explicitly here as it does not enter 
the argument in the following. 

If the l.h.s. does not vanish, i.e. if the gauge boson-quark-quark 
proper vertex $\Gamma^{\mu a}_{ij}$ acquires a pole, (\ref{WI2a}) 
shows that the quark 
self-energy $\Sigma(p)$ must possess a symmetry violating 
part resulting in a (dynamically generated) nonvanishing mass 
matrix. Now $\Gamma^{\mu a}_{ij}$ will acquire a pole if 
an effective vertex 
\beq 
\begin{minipage}{155mm}
{\hfill\epsfysize=2.5cm\epsfbox{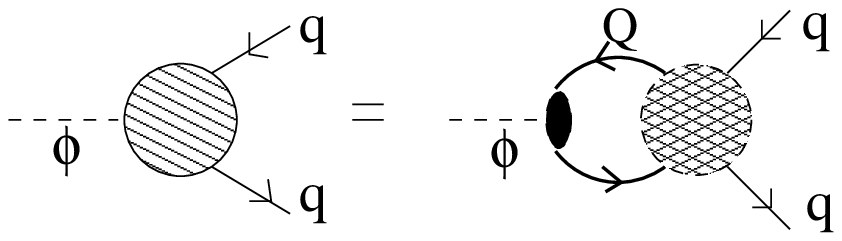}\hfill}
\end{minipage} \label{effI}
\eeq
linking the Golstone bosons to the quarks can be 
dynamically generated. 
Since communication with the techniquark sector  
occurs only through the exchange of gauge bosons, it is clear that,  
starting with massless bare quarks, this cannot happen to any finite 
perturbative order. But nonvanishing contributions to (\ref{effI}) 
can arise in the presence of mass fluctuations, i.e. massive 
quark propagators, and the question is whether a consistent 
nonperturbative solution fixing a nonzero mass can exist.

\begin{figure}[htb]
{\hfill\epsfysize=2cm\epsfbox{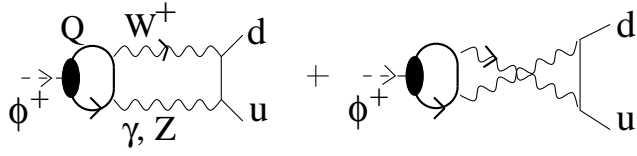}\hfill}
\caption[2ex]{\label{tc2ex}Contribution to (\ref{effI}) in presence of 
mass fluctuations.} 
\end{figure}

To lowest order in the electroweak couplings, such nonvanishing 
contributions are shown in figure \ref{tc2ex}.  
The computation of graphs in figure \ref{tc2ex}  
and in what follows is done as follows.  
Propagators for internal gauge boson lines are in Landau gauge, 
and, correspondingly, all external lines are 
transverse except for those external lines taken with longitudinal 
polarization as part of the statement of a WI (incoming line 
of momentum $q$ in the above cases). As always in dynamical symmetry 
breaking computations, the use of the Landau gauge is practically   
mandatory. It automatically 
ensures that the Goldstone pole remains massless, 
and it contributes only at a vertex to which 
the particular longitudinal leg(s) specified in the WI is (are)  attached - 
the transverse Landau propagator automatically eliminating graphs with  
longitudinal Goldstone pole contribution in all other vertices. 
All self-energies are assumed resumed giving dressed propagators.

Connecting the $\phi$'s in figure \ref{tc2ex} to incoming 
$W^+$ ($W^-$) (cp. figure \ref{tcvacp}), one obtains, for 
quark momentum $p\to 0$, a contribution to the vertex 
$\Gamma^{\mu a}_{ij}$ given by:   
\bea
 \mp\, ig\,{q^\mu \over q^2 + ie}\; 3e^2\;
\Bigg[\, J(M, m_W, m_V, m_u) \; 
q_1 m_u \;\frac{1}{2}(1\mp \gamma^5) & +&  \nonumber \\
 \qquad \qquad\qquad  J(M, m_W, m_V, m_d) \;q_2 m_d \;
\frac{1}{2}(1\pm \gamma^5)\, \Bigg]\;{\tau^{\pm}\over 2} & & \qquad 
.  \label{WI2b}
\eea 
Upper (lower) signs refer to incoming $W^+$($W^-$), and  
\bea 
J(M, m_W, 0, m) & = & {1\over 16\pi^2}\,\Bigg[\;\ln({m^2_W\over
m^2}) + O( g^2, {m^2_W\over M^2}, m\ln m)\;\Bigg] \label{J1}\\
J(M, m_W, m_Z, m) & = & {1\over 16\pi^2}\,\Bigg[\;
({m^2_W\over m_z^2}) + 
O( g^2, {m^2_W\over M^2}, m\ln m)\;\Bigg]\;.\label{J2}
\eea 
In (\ref{WI2b}) $m_V=0$ or $m_Z$ for $V=\gamma$ or $Z$, resp.; for $V=Z$ one 
must also replace $e^2 \to e^2\,\tan^2\theta_W$. Also, $
q_u\;q_d$ denote the elm charges of the quark or lepton doublet 
$(u,d)$. In obtaining (\ref{WI2b}) we made use of the GT relation, 
and approximated the fermion self-energy by 
\beq 
\Sigma (k) \approx \Sigma(0) = {\bf m}_q = 
{\rm diag}(\,m_u \;m_d\,) \label{massmtrx}\; . 
\eeq   

Consider then (\ref{WI2a}) at $p=0$. 
Inserting (\ref{WI2b}) in the l.h.s. of (\ref{WI2a}), one sees that, 
as expected, the WI is (trivially) satisfied by the 
perturbative solution ${\bf m}=0$. There is, however, also 
the possibility of a nonperturbative 
solution ${\bf m}\neq 0$ if 
\bea
3e^2\;\sum_{V=\gamma,Z} J(M, m_W, m_V, m_u) \; q_u &=& 1 \label{c1}\\
3e^2\;\sum_{V=\gamma,Z} J(M, m_W, m_V, m_u) \; q_d &=& -1 \label{c2}\;. 
\eea   

Then (\ref{c1}), (\ref{c2}) give 
\beq 
m_q = m_W\;\exp \Big\{\,-{8\pi^2\over e^2}{1\over 3|q_q|} + {1\over2}
({m_W^2\over m_Z^2})\,\tan^2\theta_W) \,\Big\}\;\Big(1 + 
O(g^2, {m^2_W\over M^2})\Big)\,, \qquad q=u,d \label{m}
\eeq 
with $q_u>0$, $q_d <0$.  

It should be noted that the loop involving the gauge bosons is 
UV convergent and receives little contribution from momenta well above 
$m_W$. In particular, in the case of photon exchange practically the 
entire contribution to the integral, the logarithmic term in (\ref{J1}),  
comes from the IR regime well below $m_W$. In fact, the integral 
becomes singular in the limit $m_q\to 0$. This justifies the approximation 
(\ref{massmtrx}). Thus, (\ref{WI2a}) is solved by a dynamical fermion 
$\Sigma(k)$ taken to be a slowing varying function representing a 
soft mass $\sim m_q$ given by (\ref{m}) for momenta well below $M$, and 
falling off rapidly in the UV region $\stackrel{>}{\sim} M$ - one is  
essentially approximating $\Sigma(k)$ by a step function.  
In a more refined approximation the falloff is fixed by the correction 
terms in (\ref{J1}), (\ref{J2}). (Analogous remarks   
apply of course also to the dynamically generated $m_W$, $m_Z$, 
and techniquark masses.)

The possible existence of such a nonperturbative solution 
to the WI would signify that the 
$m_q=0$ solution is unstable under any nonzero mass fluctuation. 
This situation occurs in a wide range of mass generation phenomena, 
including chiral symmetry breaking in QCD. 
We noted that there is no smooth $m\to 0$ limit in the conditions  
(\ref{c1})-(\ref{c2}). Also, the nonperturbative 
form  $\sim \exp(-{\rm const.}/e^2)$   
for the resulting mass ratio is characteristic of mass 
generation driven by gauge interactions as opposed to   
quadratically divergent scalar  
or four-Fermi interactions. 

By the same token, however, multiple gauge boson 
exchanges beyond the lowest two gauge boson exchange of figure 
\ref{tc2ex} cannot, in general, be ignored. 
This is because an $n+1$ gauge 
boson exchange contribution to (\ref{effI}), for example, 
may in general contain $\Big(\ln({m^2_W/m^2})\Big)^n$ terms which, 
for $m$ of the form (\ref{m}), are comparable to the 
lowest order two-rung exchange of figure \ref{tc2ex}. Thus the 
higher loop contributions cannot be ignored, and the 
solution (\ref{m}) cannot be trusted. 

This is in fact illustrated by turning to 
the remaining WI's (\ref{WI2a}), i.e. those for 
the generators $t^3$, $t^0$. 
The latter, corresponding to incoming photon, 
is trivially satisfied: due to the nonchiral photon 
coupling, contributions to both sides of (\ref{WI2a}) (in 
particular pole contributions such as those of arising from 
figure \ref{tc2ex}), vanish identically. In the case of 
incoming $Z$ connecting to $\phi^0$, however, it is not hard to 
check  that there is no nonvanishing contribution to the proper 
$\phi^0\,\bar{u}u$ and $\phi^0 \bar{d}d$ vertices 
analogous to figure \ref{tc2ex}, i.e. involving the exchange of 
just two gauge bosons.    
(This in fact reflects the structure of the electroweak 
gauge boson mass matrix.) 
As can be explicitly verified, nonvanishing such contributions 
to (\ref{effI}) begin with the exchange of 
three gauge bosons.    

Though (\ref{m}) cannot be trusted, the computation leading to it 
does show how fermion mass generation mediated by the generation of  
an effective coupling (\ref{effI}) between technipions and quarks  
is in principle possible and consistent with gauge invariance. 
The problem is how to estimate the effective vertex (\ref{effI}). 
Contributions to the blob on the r.h.s. in (\ref{effI}) 
involve  not only direct exchange of gauge bosons, 
but also technistrong interaction effects   
through, e.g., further Golstone boson exchanges. Thus a nonvanishing 
contribution of the type of figure \ref{tc2ex} 
gives rise to further processes of the type:   
\begin{figure}[h]
{\hfill\epsfysize=2.5cm\epsfbox{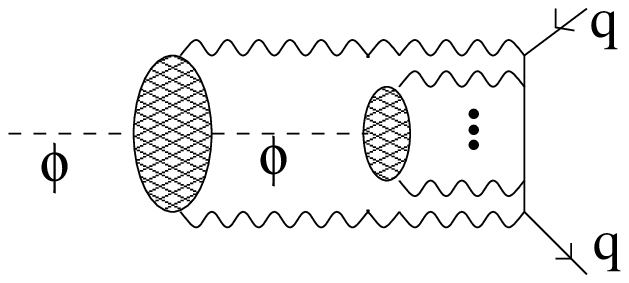}\hfill} 
%\caption[2ex]{\label{tcmex}} 
\vspace {-3mm}
\end{figure} 

\noindent These, assuming a solution for nonzero mass of the form (\ref{m}), 
can indeed give a larger contribution than other direct 
gauge boson exchange processes, and in fact further generate  
an infinite set of such exchanges by iteration.  
As always in DSB one cannot expect to be able to 
express a nonperturbatively generated effective vertex by any finite 
set of particular contributions. Rather,    
one has to determine the effective  
vertex, in this case (\ref{effI}), through self-consistent equations 
which implicitly incorporate infinite sets of graphs 
without double-counting. 

\section{Self-consistent equations for dynamical mass 
generation} 

The only systematic way of obtaining such equations is through 
the construction of the effective action for composite operators \cite{CJT} 
here applied to $\phi q\bar{q}$.   
This is done as follows. One introduces a source $v^a_{ij}(x,\bar{y},y)$ 
coupled to $\phi^a(x)\bar{q}_i(\bar{y}) q_j(y)$ in the functional 
integral for techniquarks, electroweak gauge bosons and quarks 
to obtain the generating functional $iW[v] = \ln Z[v]$. 
(Note that $\phi$ is itself a composite field, but we assume that the 
techniquark sector is adequately described by the effective linear 
sigma model description, cp. (\ref{gold}).) 
The effective action $\Gamma$ for the 3-point vertex 
\beq
g^a_{ij}= {\delta W[v]\over \delta v^a_{ij}} \equiv 
\Delta^{ab}S_{ik} \,\gamma^b_{kl}\, S_{lj} \label{effver}
\eeq 
is then defined by 
\beq 
\Gamma[\gamma] = W[v] - v\cdot g \;,\qquad 
{\delta \Gamma\over \delta g } = -v \;. \label{effact}
\eeq 
Here $S$, $\Delta$ denote the full quark and Goldstone 
propagators, and hence $\gamma$ represents the proper 
vertex (\ref{effI}). The effective action $\Gamma[\gamma]$     
can be expanded in the form:  
\beq 
\Gamma[\gamma] = \Gamma_0 - \frac{1}{2}\,
\gamma^b_{lk}\,S_{jl}\,\Delta_{ba}\,S_{ki}\,
\gamma^a_{ij} + \Gamma_3[\gamma] \,,\label{effexp1} 
\eeq 
where 
\beq 
\Gamma_3 = \Bigg\{ \mbox{sum of all only trivially 3PR 
vacuum graphs} \Bigg\}.\label{effexp2}
\eeq  
(Obvious condensed notation with summation-integration 
over repeated generalized indices is used.) An only  trivially 
3PR (3-particle-reducible) graph is a 2PI (2-particle irreducible) 
graph that can be cut into two parts by cutting three lines 
if and only if one and only one of the two parts 
consists of a single 3-point vertex. 
Otherwise the graph is 3PI. All graphs are computed with 
dressed propagators.

The $\gamma^a_{ij}$ vertices are determined by 
the minimization of the effective action (\ref{effact}): 
\beq
{\delta \Gamma[\gamma]\over \delta \gamma} = 0 . \label{min} 
\eeq 
Imposing (\ref{min}) gives, to lowest order in the skeleton loop 
expansion (\ref{effexp2}), the following coupled set of 
self-consistent equations: 
\beq
\begin{minipage}{155mm}
\epsfysize=4.5cm\epsfbox{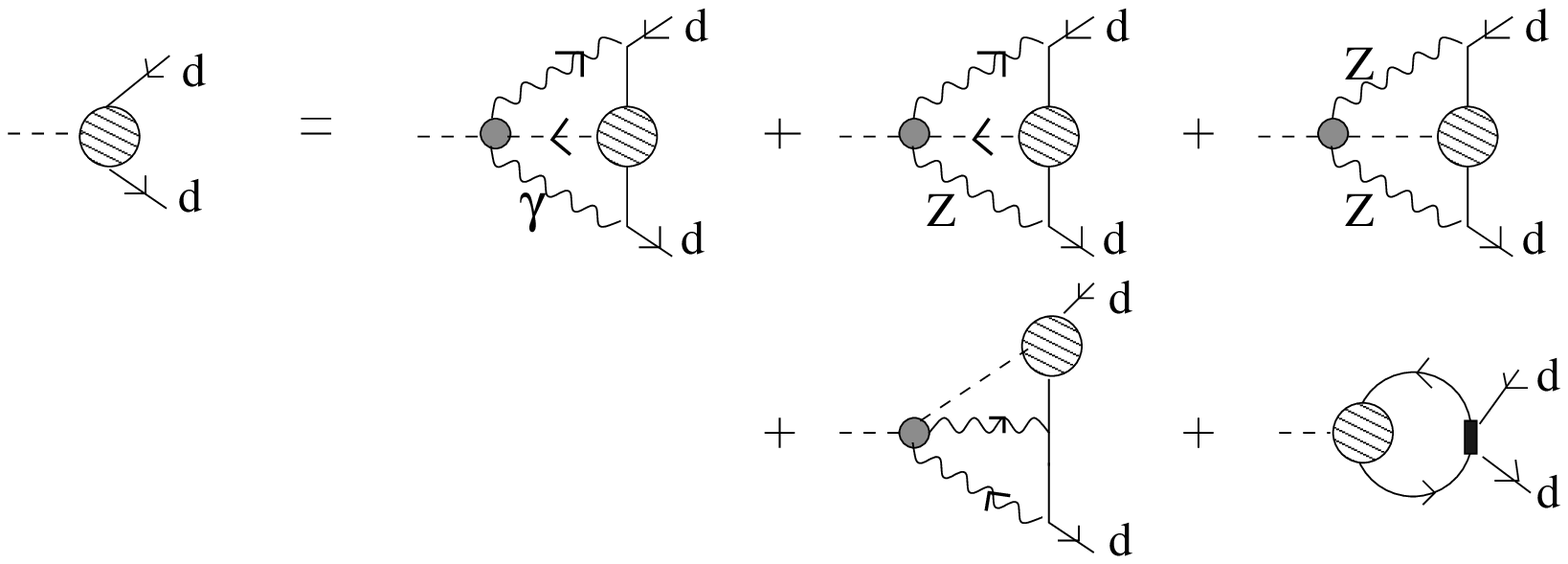}
\end{minipage} \label{sc1}
\eeq
\beq
\begin{minipage}{155mm}
\epsfysize=4.5cm\epsfbox{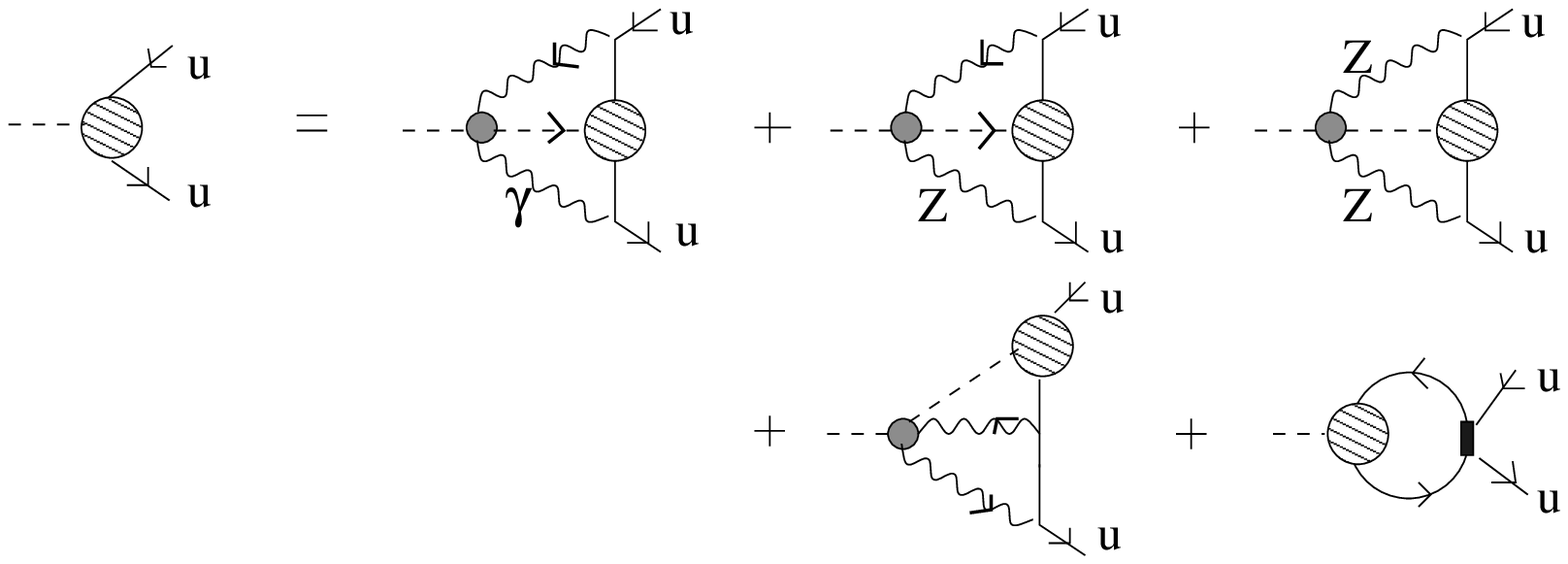}
\end{minipage} 
\label{sc2}
\eeq
There is also the corresponding set of equations with the  
$\phi^+\bar{u}\,d$, $\phi^-\bar{d}\,u$ vertices on the l.h.s. 
In (\ref{sc1})-(\ref{sc2}) oriented wavy lines represent $W^{\pm}$'s 
and oriented dashed lines represent $\phi^{\pm}$'s, with 
positive charge flow into a vertex in the direction of the arrow. 
Unoriented wavy and dashed lines represent 
$\gamma$, $Z$, and $\phi^0$, respectively. 
For brevity, each graph depicted in (\ref{sc1})-(\ref{sc2}) 
is understood to stand for a set of     
graphs that, in addition to the graph in question, 
also includes all `crossed' graphs that may be  
formed by permutations of its vertices.     
The small filled blobs represent the techniquarkloop-induced 
vertices between $W^{\pm}$, $Z$, 
$\gamma$ and technipions, e.g 
\beq
\begin{minipage}{3cm}
\epsfysize=2cm\epsfbox{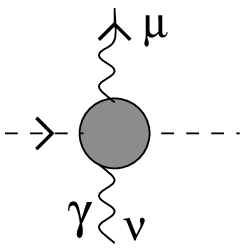}
\end{minipage}  \quad \equiv  
\begin{minipage}{3cm}
\epsfysize=2.5cm\epsfbox{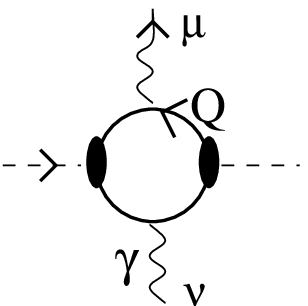}
\end{minipage} \quad 
    =  -ig^2\,s_W\,{1\over2}g^{\mu\nu}\; +\; O(k^2/M^2)\,, \label{tcv}  
\eeq
(with $s_W\equiv \sin \theta_W$, $k$ of the order of the external 
momenta). The last graph on the r.h.s. in (\ref{sc1}) and 
(\ref{sc2}) is the contribution from the 
four-quark interactions (\ref{4f}) (square vertices).

The Goldstone boson is massless (Landau gauge). 
The self energy in the dressed quark propagator in 
(\ref{sc1})-(\ref{sc2}), on the other hand, is related by the WI 
(\ref{WI2a}) to the vertex (\ref{effI}). Indeed, making use of  
the GT relation,  (\ref{WI2a}) is seen to be 
equivalent to the following relation between   
the self energy, in the approximation (\ref{massmtrx}), and 
the effective $\gamma$ vertices:     
\bea
\begin{minipage}{2cm}
\epsfysize=2cm\epsfbox{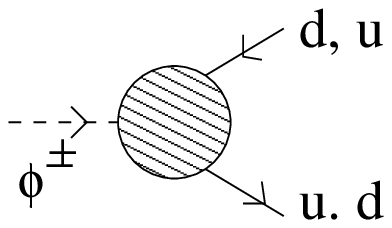}\end{minipage}&  =& \qquad
\mp {1\over\sqrt{2}}\,g\,\Bigg[\,{m_u\over m_W} 
             \;\frac{1}{2}(1\mp \gamma^5)  
              - {m_d\over m_W} \; \frac{1}{2}(1\pm \gamma^5)\, 
                    \Bigg]\;. \label{eff1} \\
\begin{minipage}{2cm}
\epsfysize=2cm\epsfbox{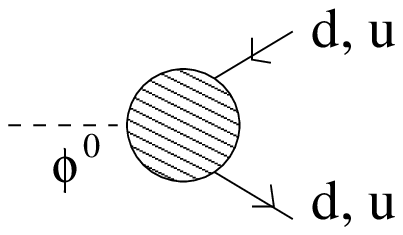}\end{minipage} &=& \qquad
{1\over2}\,g\;{{\bf m}_q\over m_W}\cdot\tau^3\,\gamma^5 
            \label{eff2}
\eea 
(In (\ref{eff1}) upper (lower) signs refer to incoming 
$d$ ($u$)).

(\ref{sc1})-(\ref{sc2}) are to be solved together with 
(\ref{eff1}), (\ref{eff2}). 
Note that these equations are indeed what one would 
basically expect to get in a self-consistent Hartree-Fock approximation 
for the vertices (\ref{eff1})-(\ref{eff2}). The effective action 
(\ref{effexp2}), however,  provides in principle 
a systematic approximation scheme. $m_u=m_d=0$ is seen to be, 
trivially, always a solution; we are looking for a nontrivial 
nonperturbative ${\bf m}_q\not=0$ solution.

Note that,  
when terms suppressed by inverse powers of $M$ are neglected, 
the techniquarkloop-induced vertices are essentially those of 
the standard model (cp. second equality in (\ref{tcv})). 
Even so, the system (\ref{sc1})-(\ref{sc2}) 
appears rather formidable due to the remaining $2$-loop 
structures and the large number of graphs involved.     

What makes computations  feasible is the fact that 
in the limit of vanishing external momenta all $2$-loop graphs 
are in principle explicitly computable in terms of Spence (di-log) 
functions \cite{dBV}. By a variety of tricks, every $2$-loop integral with 
numerator momentum tensor structures can be reduced to a series of 
$2$-loop scalar field theory integrals, which in turn can be 
related to standard types involving four propagators. These 
procedures generally generate large number of terms. The 
scalar integrals can then be explicitly integrated in terms 
of dilogs, which can finally be expanded in ratios of masses. 
In the case at hand the use of Landau gauge gives numerator 
tensor structures with up to six powers of momenta making 
computation of graphs very laborious and lengthy. The 
structure of the results, however, is rather easily stated.  

In computing the r.h.s. of (\ref{sc1})-(\ref{sc2}) we assume that  
$m_q = m_d$, $m_u$ are much smaller than all other mass scales, and are  
interested in exhibiting the expected large double and single 
pure $\ln(m_q/m_W)$ terms. Fortunately, not all graphs in 
(\ref{sc1})-(\ref{sc2}) provide such large logs. Examination of 
the expansion in mass ratios of all possible dilogs resulting 
from the various graphs shows that 
only graphs containing a photon line give such logs. It 
should be pointed out that this is 
not really a consequence of the masslessness of the photon, 
but of the (near) degeneracy of the $Z$, $W^+$, and $W^-$ masses. 
We also take the cutoff in 
(\ref{4f}) to be of the order of the technicolor CSB cutoff. 

We have computed explicitly planar graphs (taking the 
techniquark loop structure into account) on the r.h.s 
of (\ref{sc1})-(\ref{sc2}). Though this is a  
lengthy computation, the structure of the resulting two conditions 
for $m_d$, $m_u$ is simple. Taking $G_2$ in (\ref{4f}) to be 
small compared to $G_1$, $G_3$, one finds:    
\bea 
\Big({g^2\over 16\pi^2}\Big)^2\,\Bigg[\,s^2_w q_q \,\Big[\,{9\over8}   
\,\ln^2({m_W^2\over m_q^2}) + (c_1 + c^\prime_1 \ln({M^2\over m_W^2}))\,
\ln({m_W^2\over m_q^2})\Big] & & \label{scconds}\\ \nonumber
 +O\Big(\ln^2({M^2\over m_W^2}), \,1, \, 
{m^2_W\over M^2}, \,{m_{u,d}^2\over m_W^2}\ln({m^2_{u,d}\over m_W^2})
\Big)\,\Bigg]  + (\,G_1 \pm G_3\,)\,[\,1 + 
O({m_q^2\over M^2}\,\ln({M^2\over m_q^2}))\,] 
& = &\pm1 
\eea 
for $m_q=m_u$, $q_q=q_u$ and upper sign, and for $m_q=m_d$, $q_q=q_d$ 
and lower sign. $c_1$, $c_1^\prime$ are numerical constants of order unity, 
and only the order of all other resulting terms is indicated. 
(\ref{scconds}) then gives: 
\beq
m_q \sim m_W\;\exp \Big\{\,-{8\pi^2\over g^2} \bigg( {1-C_q\over  
s_W^2|q_q|c_2}\bigg)^{1/2} \,\Big\}\;\,, \qquad q=u,d \label{scm}
\eeq 
where $c_2=9/8$, and $C_u=(G_1+G_3)$, $C_d=G_1-G_3$.  

The masses obtained in (\ref{scm}) are naturally tiny compared 
to $m_W$. We have not examined the possibility of solutions 
to ({\ref{sc1})-(\ref{sc2}) with $m_q$ not much smaller than 
all other mass scales. 

A complete derivation of (\ref{scm}) would include also computation 
of the nonplanar contributions in (\ref{sc1})-(\ref{sc2}) which 
has not yet been completed. These will change the numerical value 
of the coefficient $c_2$, but should not change the qualitative 
behavior. 

\section{Discussion}

The exponential form of the solution (\ref{scm}) is 
due to the presence of the gauge interactions which are responsible  
for the pure log terms in (\ref{scconds}). The role of the 
four-fermion interactions (\ref{4f}), however, should now be noted. They 
contribute the factors $(1-C_q)$ in the exponent.  
Variations in the values of the couplings in (\ref{4f}), therefore, 
may produce very large variations in the dynamically generated masses. 
More importantly, these couplings may be adjusted to stabilize 
(\ref{scm}) against higher corrections from the expansion 
(\ref{effexp2}), i.e. render all higher powers of logs in 
higher loop contributions smaller.\footnote{A well-known    
example of this arises in the minimization of the 
Coleman-Weinberg potential.}  

It is useful to contrast this to 
mass generation driven only by four-fermion 
interactions, as in top condensation models. In that case one 
considers only the last diagram on the r.h.s. of (\ref{sc1})-(\ref{sc2}).  
Correspondingly one has to balance the $(m_q^2/M^2)\,
\ln({M^2/m_q^2})$ piece of the term   
proportional to the $G_i$'s on the l.h.s. of (\ref{scconds}) against 
a constant. A solution then requires extreme 
fine-tuning of the couplings $G_i$ (above a critical (strong) value) 
in order to obtain a small mass $m_q$ relative to the 
cutoff $M$.\footnote{This is again the usual fine-tuning 
due to quadratic divergences with a fundamental Higgs in a 
different guise.}  
In the present case, we need  not take the four-fermion couplings 
in (\ref{4f}) to be above critical to drive mass generation, and 
$m_q^2\ln m_q^2$ terms are 
unimportant as they can produce only 
tiny corrections to the leading behavior (\ref{scm}). The 
constant piece contributions in (\ref{scconds}) coming from (\ref{4f}) 
can then have the important 
effects pointed out above without any excessive fine-tuning.

We explored the possibility of the above nonperturbative 
mechanism for fermion mass generation within the familiar context of 
the standard minimal electroweak model. 
The mechanism, however, 
is quite general, and may be realized in a wider context and at 
different scales. Not all contributions to all the 
quark and lepton masses need of course arise at a single scale.   
A natural application for the proposed mechanism is in models where 
the electroweak gauge bosons are replaced by a system of 
gauge bosons with some nondegenerate masses (cp. remarks preceding 
(\ref{scconds})) of a much higher, perhaps 
unification, scale. The analog of (\ref{scm}) then  
produces very large natural hierarchies.    
One such simple model will be treated elsewhere. 

\vspace{.5cm}
I would like to thank Z. Bern and A. Grant for discussions.

\end{document}